\documentclass[]{mn2e}
\usepackage{graphicx}
\usepackage{epsfig}
\usepackage{amsmath}
\newcommand       \Angstrom     {\,{\rm \AA}}

\newcommand       \mm           {\,{\rm mm}}

\newcommand       \K            {\,{\rm K}}

\newcommand	  \yr		{\,{\rm yr}}

\newcommand     \gtsim  {\lower.5ex\hbox{$\buildrel > \over \sim$}}
\newcommand     \ltsim  {\lower.5ex\hbox{$\buildrel < \over \sim$}}
\newcommand     \simgt  {\lower.5ex\hbox{$\buildrel > \over \sim$}}
\newcommand     \simlt  {\lower.5ex\hbox{$\buildrel < \over \sim$}}

\newcommand       \mum          {\,{\rm \mu m}}

\newcommand       \Msun         {\,{M_\odot}}

\newcommand       \simali       {\sim\,}
\newcommand       \rhom         {\rho_{\rm m}}
\newcommand       \rhoR         {\rho_{\rm R}}

\newcommand	  \md	        {m_{\rm dust}}
\newcommand	  \Vd	        {V_{\rm dust}}
\newcommand	  \mmin	        {m_{\rm dust}^{\rm min}}
\newcommand	  \cext	        {C_{\rm ext}}
\newcommand	  \cabs	        {C_{\rm abs}}
\newcommand	  \csca  	{C_{\rm sca}}

\title{On the absorption properties of metallic needles}
\author[Xiao, Li, Li \& Chen]
       {C.Y.~Xiao$^{1}$\thanks{xiaocunying@bnu.edu.cn},
        Qi~Li$^{1,2}$\thanks{201531160001@mail.bnu.edu.cn},
        Aigen~Li$^{2}$\thanks{lia@missouri.edu}, and
        J.H.~Chen$^{3}$\thanks{jhchen@hunnu.edu.cn}\\
        $^1$Department of Astronomy,
                Beijing Normal University,
                Beijing 100875, China\\
        $^2$Department of Physics and Astronomy,
             University of Missouri,
             Columbia, MO 65211, USA\\
        $^3$Department of Physics, 
               Hunan Normal University, 
               Changsha, Hunan 410081, China\\
             }
\begin{document}
\date{}
\pagerange{\pageref{firstpage}--\pageref{lastpage}} \pubyear{2020}

\maketitle

\label{firstpage}
\begin{abstract}
Needle-like metallic particles have been suggested
to explain a wide variety of astrophysical phenomena,
ranging from the mid-infrared interstellar extinction
to the thermalization of starlight to generate 
the cosmic microwave background.
These suggestions rely on the amplitude 
and the wavelength dependence of 
the absorption cross sections of metallic needles.
On the absence of an exact solution to the absorption 
properties of metallic needles,
their absorption cross sections are often derived 
from the antenna approximation. 
However, it is shown here that the antenna approximation
is not an appropriate representation 
since it violates the Kramers-Kronig relation.
Stimulated by the recent discovery of 
iron whiskers in asteroid Itokawa
and graphite whiskers in carbonaceous chondrites,
%and magnetite whiskers in Martian meteorites,
we call for rigorous calculations of the absorption
cross sections of metallic needle-like particles,
presumably with the discrete dipole approximation. 
%{\bf 
We also call for experimental studies 
of the formation and growth mechanisms 
of metallic needle-like particles 
as well as experimental measurements 
of the absorption cross sections 
of metallic needles of various aspect ratios 
over a wide wavelength range
to bound theoretical calculations.
%
%}
%
\end{abstract}
\begin{keywords}
dust, extinction -- infrared: ISM -- intergalactic medium
%-- microwave background
\end{keywords}

\section{Introduction}\label{sec:intro}
In the Galactic interstellar medium (ISM),
typically 90\% or more of the iron (Fe) element 
is missing from the gas phase (Jenkins 2009), 
        suggesting that Fe is the largest elemental
        contributor to the interstellar dust mass 
        after O and C and accounts for $\simali$25\% 
        of the dust mass in diffuse interstellar regions.
        However, as yet we know little about the nature 
        of the Fe-containing material.
        Silicate grains provide a possible reservoir 
        for the Fe in the form of interstellar pyroxene
        (Mg$_x$Fe$_{1-x}$SiO$_3$) 
        or olivine (Mg$_{2x}$Fe$_{2-2x}$SiO$_4$) analogues. 
        %
        %Iron abundances and depletions in the ISM 
        %often diverge from the pattern shown by Si and Mg. 
        %This is evidence that Fe is not tied to the same grains 
        %as silicon, and therefore that most silicate grains 
        %are likely magnesium based. 
        %The presence of iron depletion in the SMC then 
        %suggests that this element is probably incorporated  
        %into grain types such as metals or oxides...  
        %
Nevertheless, the shape and strength of 
the 9.7\,$\mu$m interstellar silicate feature 
in extinction suggest that the silicate material 
is Mg-rich rather than Fe-rich (Poteet et al.\ 2015)
and therefore that a substantial fraction ($\simali$70\%) 
of the interstellar Fe is in other forms 
such as iron oxides (e.g., Fe$_3$O$_4$),
iron sulfides, or metallic iron
(see Draine \& Hensley 2013, 
Hensley \& Draine 2017, 
Westphal et al.\ 2019).
Also, Fe abundances and depletions in the ISM 
often diverge from the pattern shown by Si and Mg,
suggesting that Fe is not tied to the same grains 
as Si and Mg, and therefore that most silicate grains 
are likely Mg-based while Fe is probably incorporated  
into grain types such as metals or oxides.  

As a matter of fact,
small metallic particles were among the materials initially 
proposed to be responsible for the interstellar reddening
(Schal\'{e}n 1936, Greenstein 1938).
About a half century later, the idea of metallic grains as an 
interstellar dust component was reconsidered by Chlewicki \& 
Laureijs (1988) who argued that small iron particles
with a typical size of $\simali$70$\Angstrom$
would obtain an equilibrium temperatures of 
$\simali$53$\K$ in the Galactic cirrus diffuse ISM 
and account for the 60$\mum$ emission 
measured by the {\it Infrared Astronomical Satellite} 
(IRAS) broadband photometry. 
%
%{\bf 
Iron grains have long been neglected
since it is commonly accepted that 
$\simali$2/3 of the IRAS 60$\mum$ emission
of the Galactic cirrus arises from stochastically
heated nano-sized dust grains (Draine \& Li 2001)
and $\simali$1/3 of that arises from ``classical'', 
submicron-sized grains which attain equilibrium 
temperatures in the ISM (Li \& Draine 2001).
%
%}
%

Thanks to the successful operations of
the {\it Infrared Space Observatory} (ISO)
and the {\it Spitzer Space Telescope},
numerous observations have shown that
the mid infrared (IR) extinction 
at $3\mum <\lambda< 8\mum$
is flat or ``gray'' for both diffuse
and dense environments
(Lutz 1999, 
Indebetouw et al.\ 2005,
Jiang et al.\ 2006, 
Flaherty et al.\ 2007,
Gao et al.\ 2009,
Nishiyama et al.\ 2009, 
Wang et al.\ 2013,
Xue et al.\ 2016,
Hensley \& Draine 2020).
The flat mid-IR extinction, 
which is anomalous 
compared to the deep minimum expected 
from standard interstellar dust models 
(see Draine 1989, Weingartner \& Draine 2001),
has been suggested as resulting from
very large, micron-sized 
graphitic grains (Wang et al.\ 2015a) 
or ice grains (Wang et al.\ 2015b).
Alternatively, needle-like metallic grains
%with a typical length ($l$) to radius ($a$) 
%ratio of $l/a\approx600$ 
were suggested by Dwek (2004a) 
as an explanation for the flat mid-IR extinction.
Metallic needles 
%with an $l/a$ ratio of about 3000 
were also suggested by Dwek (2004b) as the source
for the submillimeter excess observed along 
the line of sight to the Cas~A supernova remnant (SNR)
by Dunne et al.\ (2003) 
since highly elongated needles are powerful emitters
in the submillimeter and millimeter wavelengths
(see Edmunds \& Wickramasinghe 1975).\footnote{% 
  %If attributed to regular interstellar dust that formed in  
  %the SN ejecta, the submillimeter emission would have 
  %implied a dust mass exceeding the mass of refractory 
  %elements that could have formed in supernova explosion. 
  %A needle origin for the submillimeter emission 
  %would require only a small fraction of the metallic iron 
  %synthesized in the explosion to have formed 
  %in the ensuing expansion (Dwek 2004a). 
  However, the need for needles 
  has since been alleviated with subsequent observation 
  and analysis that showed that the submillimeter emission 
  emanates not from the remnant, but from a molecular cloud 
  along the line of sight 
  (Krause et al.\ 2004, Wilson \& Batrla 2004).
  Gomez et al.\ (2005) applied the iron needle model to 
  the submillimetre emission from the Kepler SNR 
  to which the contribution from foreground clouds 
  is negligible (Gomze et al.\ 2009) and found  that
  the iron needle model is not very favorable.
  }

Extraterrestrial metallic needle-like particles 
are actually not as exotic as they appear. 
Nanometer-sized magnetite (Fe$_3$O$_4$) whiskers 
%associated with carbonates in fracture zones 
%within Martian meteorite ALH84001 
were seen in Martian meteorite ALH84001 
with analytical transmission electron microscopy
(Bradley et al.\ 1996).\footnote{%
%   {\bf
   We should note that those magnetite whiskers 
   identified in Martian meteorite ALH84001 
   formed in a hot, hydrous, planetary crust 
   and not in the solar nebula
   and therefore would not constitute
   an observable astrophysical phenomenon.
%   }
   }
Graphite whiskers were discovered in 
carbonaceous chondrites 
via Raman imaging and electron microscopy
(Fries \& Steel 2008).
Very recently, Matsumoto et al.\ (2020) reported
the discovery of iron whiskers on particles 
from the asteroid samples returned by 
the Japanese Hayabusa mission to asteroid Itokawa.
This discovery was based on high-resolution images 
of Itokawa dust particles obtained with 
a transmission electron microscope.
%{\bf
The identification of whiskers in extraterrestrial samples
has led to renewed interest in metallic needles.
%}
%

Historically, 
exceedingly elongated metallic needles, 
%with a length ($l$) over radius ($a$) ratio $l/a \approx 10^5$, 
presumably present in the intergalactic medium (IGM), 
have been suggested  by Hoyle \& Wickramasinghe (1988) 
as a source of starlight opacity, 
creating a non-cosmological microwave background by the thermalization 
of starlight in a steady-state cosmology. 
Wright (1982) and Aguirre (2000) examined the possibility 
that needles could have thermalized the light from stars 
formed after a cold big bang and generated the microwave 
background in cold big bang model. Needles have also been 
suggested by Aguirre (1999) and Banerjee et al.\ (2000) 
as a source of the gray opacity needed to explain 
the observed redshift-magnitude relation of Type Ia
supernovae (Riess et al.\ 1998; Perlmutter et al.\ 1999) 
without resorting to a positive cosmological constant. 
Narlikar et al.\ (2003) invoked needles to
explain the microwave background anisotropy detected by 
the Wilkinson Microwave Anisotropy Probe (Bennett et al.\ 2003) 
in a steady state cosmology. 
Very recently, the role of dust 
in the rethermalization of the cosmological 
microwave background photons
has been thoroughly explored by Melia (2020).

Apparently, an appropriate evaluation of 
the absorption properties as a function of wavelength 
for metallic needles is critical in assessing the role 
of needles in explaining a wide variety of 
astrophysical phenomena. 
%In particular, to study  their ability to thermalize 
%the CMB (Wright 1982; Hoyle \& Wickramasinghe 1988),
%their contribution to the IR interstellar extinction 
%(e.g. see Dwek 2004a), and the dust production rate in 
%supernova explosion through constraining the dust mass 
%from the submillimeter emission of metallic needles 
%(Dwek 2004b), a detailed knowledge of the mass absorption 
%coefficient in the wavelength range from the far ultraviolet 
%(UV) to millimeter is required.
%
Unfortunately, there exists no exact solution to 
the interaction of metallic needles with electromagnetic
radiation. In the literature, their absorption properties 
are often derived either from needle-like spheroids
under the Rayleigh approximation assumption 
(see Li 2003a and references therein) 
or from the antenna approximation (Wright 1982).
However, we have shown in a previous paper (Li 2003a) 
that the Rayleigh approximation is invalid 
since the Rayleigh criterion is not satisfied 
for highly conducting needles. 
We will show in this work that the antenna approximation 
is not applicable either since it violates the Kramers-Kronig 
relation, a fundamental physical principle.
We stress that our goal here is not to directly 
assess the role of metallic needles in explaining
any specific astrophysical phenomenon,
but rather to assess the applicability 
of the antenna approximation for
calculating the absorption cross sections 
of metallic needles which serve as an essential
ground for the former.

\section{Antenna Approximation for Metallic Needles}\label{sec:antenna}
Following Wright (1982), we consider a conducting 
needle-like dust particle represented by a circular cylindrical 
of radius $a$, length $l$ ($a\ll l$), mass density $\rhom$, 
and mass $\md = \pi a^2l\rhom$. Let $\rhoR$ be its resistivity, 
$R =\rhoR l/\pi a^2$ be its resistance,
and $V$ be the voltage on the needle.  
If the resistance is the primary limit to current flow,
then the power $P$ absorbed in the grain is
\begin{equation}\label{eq:power}
P = \frac{V^2}{R} = \frac{1}{3} \frac{\left(El\right)^2}{R}
= \frac{1}{3} \frac{\pi a^2 E^2 l}{\rhoR} ~~,
\end{equation}
where $E$ is the amplitude of the incident electric field,
and the factor of $1/3$ arises from averaging over 
the angles of incidence between the electric field and 
the needle (Wright 1982).
The absorption cross section is given by 
\begin{equation}\label{eq:cabs1}
C_{\rm abs} = P/S = \frac{4\pi}{3c}\frac{\pi a^2 l}{\rhoR} ~~,
\end{equation}
where $S = \left(c/4\pi\right)E^2$ is 
the time-averaged Poynting vector of the incident radiation,
and $c$ is the speed of light.
The long wavelength cutoff for $C_{\rm abs}$ will come
when the capacitive reactance of the needle equals its
resistance. This gives a long wavelength cutoff 
(see Wright 1982)
\begin{equation}\label{eq:lambda0}
\lambda_0 = \frac{1}{2}\rhoR c 
\frac{\left(l/a\right)^2}{\ln\left(l/a\right)} ~~.
\end{equation}
Therefore, the absorption cross section for a conducting
metallic needle is given by
%
%\begin{equation}\label{eq:cabs2}
%C_{\rm abs}(\lambda) = 
%\left\{\begin{array}{lr}
%\frac{4\pi}{3c}\frac{\pi a^2 l}{\rhoR}~, &\lambda \le \lambda_0~,\\
%\frac{4\pi}{3c}\frac{\pi a^2 l}{\rhoR} \left(\lambda/\lambda_0\right)^{-2}~, &\lambda > \lambda_0~.\\
%\end{array}\right.
%\end{equation}
%
%
\begin{eqnarray}\label{eq:cabs2}
C_{\rm abs}(\lambda) & = &
%\left\{\begin{array}{lr}
\frac{4\pi}{3c}\frac{\pi a^2 l}{\rhoR}~~,~~\lambda \le \lambda_0~,\\
&=&\frac{4\pi}{3c}\frac{\pi a^2 l}{\rhoR} \left(\lambda/\lambda_0\right)^{-2}~~,~~\lambda > \lambda_0~.
\end{eqnarray}
%
%

%%% Figure 1 %%%
\begin{figure*}
\centering
\includegraphics[width=.5\textwidth]{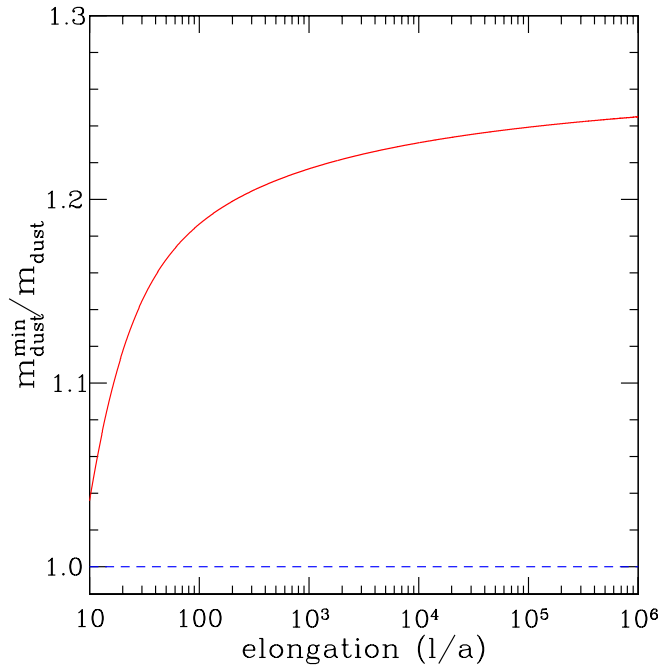}
%\vspace{2mm}
\caption{
        \label{fig:mkk}
        The ratio of the Kramers-Kronig lower bound
        on the metallic needle mass $\mmin$ to the real
        needle mass $\md$. For highly conducting needles
        (the static dielectric constant 
         $\varepsilon_0\rightarrow \infty$),
        this ratio [$\mmin/\md = \left(4/9\pi F\right) 
        \left(l/a\right)^2/\ln\left(l/a\right)$]
        depends only on the elongation of the needle,
        regardless of the physical and chemical properties 
        of the needle material (e.g., the resistivity $\rhoR$).
        The dimensionless factor $F$ for elongated conducting
        dust is calculated from eqs.\,\ref{eq:F}--\ref{eq:e}.
        %Since $\mmin/\md >1$ for all elongation $l/a$,
        %the antenna theory is not an appropriate 
        %representation for the absorption cross sections of
        %metallic needles. 
        }
\end{figure*}

\section{Constraints from the Kramers-Kronig Relations}\label{sec:kk}
The antenna representation of the absorption cross
sections for metallic needles (see eq.\,\ref{eq:cabs2}) is
straightforward and easy to compute. But it is unclear whether
it is appropriate for extremely elongated conducting needles.
In this section, we will test this approximation in terms of 
the Kramers-Kronig relations.  

Let $\cext(\lambda)$ be the extinction cross sections
of a conducting needle-like particle at wavelength $\lambda$, 
and $\int_{0}^{\infty} \cext(\lambda) d\lambda$ be 
the extinction integrated over the entire wavelength 
range from 0 to $\infty$. As shown by Purcell (1969),
the Kramers-Kronig dispersion relations can be used to 
relate $\int_{0}^{\infty} \cext(\lambda) d\lambda$
to the grain volume $\Vd$ through
\begin{equation}\label{eq:kk}
\int_{0}^{\infty} \cext(\lambda) d\lambda 
= 3 \pi^2 \Vd F\left(\varepsilon_0;{\rm shape}\right) ~~,
\end{equation}
where $F$ is the orientationally-averaged polarizability 
relative to the polarizability of an equal-volume sphere 
(Purcell 1969). The dimensionless factor $F$ depends only 
upon the grain shape and the static (zero-frequency) 
dielectric constant $\varepsilon_0$ of the grain material 
(Purcell 1969, Draine 2003, Li 2003b).
For prolates of axial ratio $l/a$,
where $l$ is the semiaxis along the symmetry axis, 
and $a$ is the semiaxis perpendicular to the symmetry axis, 
$F$ is given by
\begin{equation}\label{eq:F}
F\left(\varepsilon_0;l/a\right) 
= \frac{\varepsilon_0-1}{9}
\left[\frac{1}{1+L_{\|}\left(\varepsilon_0-1\right)}
+\frac{2}{1+L_{\bot}\left(\varepsilon_0-1\right)}\right] ~~,
\end{equation}
where $L_{\|}$ and $L_{\bot}$ are the depolarization factors 
parallel and perpendicular, respectively, to the grain symmetry 
axis and are given by 
\begin{equation}\label{eq:Lpara}
L_{\|} = \frac{1-e^2}{e^2} 
      \left[\frac{1}{2e} {\rm ln}\left(\frac{1+e}
      {1-e}\right)-1\right] ~~,
\end{equation}
\begin{equation}
L_\bot = \frac{1-L_{\|}}{2} ~~,
\end{equation}
where the eccentricity $e$ is
\begin{equation}\label{eq:e}
e = \sqrt{1-(a/l)^2} ~~.
\end{equation}
The $F$ factors for highly conducting 
($\varepsilon_0\rightarrow \infty$) prolates and oblates 
have been calculated in Li (2003b). 
For extremely elongated prolates ($l\gg a$),
$L_{\|}\approx \left(a/l\right)^2 \ln\left(l/a\right)$
and therefore 
$F(\varepsilon_0;l/a)\approx 
\left(l/a\right)^2/\left[9\ln\left(l/a\right)\right]$
for highly conducting ($\varepsilon_0\rightarrow \infty$)
{\it and} extremely elongated prolates.
We will take the $F$ factor of metallic needles
to be that of highly conducting, extremely elongated prolates.
%

%For a given set of absorption cross sections as a function
%of wavelength, we can apply Eq.(\ref{eq:kk}) to the interstellar
%space, the intergalactic space, or circumstellar envelopes
%to obtain a lower bound on the needle mass $\mmin$,
%taking the interstellar space, the intergalactic space, or
%circumstellar envelopes to be a vacuum sparsely populated 
%by conducting needles.
%Since $\cext = \left(\cabs + \csca\right) > \cabs >0$
%(where $\cabs$ and $\csca$ are the absorption and scattering 
%cross sections, respectively; and both are positive numbers),
%the integration of $\cabs$ over $\lambda=0$ to $\infty$ 
%therefore represents a lower limit to 
%$\int_{0}^{\infty} \cext(\lambda) d\lambda$, 
%and implies a lower limit to the volume of space which 
%must be filled by needle particles
%and hence a lower limit to the needle mass.

For a given set of absorption cross sections as a function
of wavelength, we can apply eq.\,\ref{eq:kk} to obtain a lower 
bound on the needle mass $\mmin$. 
Since $\cext = \left(\cabs + \csca\right) > \cabs >0$
(where $\cabs$ and $\csca$ are the absorption and scattering 
cross sections, respectively; and both are positive numbers),
the integration of $\cabs$ over $\lambda=0$ to $\infty$ 
therefore represents a lower limit to 
$\int_{0}^{\infty} \cext(\lambda) d\lambda$, 
and implies a lower limit to the needle mass
\begin{equation}\label{eq:mmin}
\mmin = \frac{\rhom}{3\pi^2 F\left(\varepsilon_0;l/a\right)}
\int_{0}^{\infty} \cabs(\lambda) d\lambda 
= \frac{8\lambda_0 \rhom}{9\pi c F} \frac{\pi a^2 l}{\rhoR} ~~,
\end{equation}
where we have used eq.\,\ref{eq:cabs2}, the antenna 
representation of the absorption cross sections 
for conducting needles. 
Therefore, we can obtain the ratio of $\mmin$,
a lower limit to the needle mass, 
to $\md$, the actual needle mass,
\begin{equation}\label{eq:mmin2md}
\frac{\mmin}{\md} = \frac{8\lambda_0}{9\pi c F \rhoR} 
= \frac{4}{9\pi F} 
\frac{\left(l/a\right)^2}{\ln\left(l/a\right)} ~~.
\end{equation} 
Since $\mmin$ is a lower limit to $\md$, we should expect
$\mmin/\md <1$ for any physically reasonable systems.
However, as shown in Figure \ref{fig:mkk}, we find 
$\mmin/\md >1$ for needles of both small and 
large elongation $l/a$ (the ratio of the length $l$ 
to the radius $a$), implying that the antenna approximation 
shown in eqs.\,\ref{eq:lambda0},\ref{eq:cabs2}
is not an appropriate representation for the absorption 
cross sections of metallic needles. 

It is interesting to note that since $\mmin/\md$ is independent
of the resistivity $\rhoR$ of the needle material (provided that
the needle is made of conducting materials), our conclusion
is applicable to any conducting needles, 
e.g, graphite whiskers, iron needles, nichrome (Ni-Cr) needles,
and iron needles containing a small fraction of embedded impurities.
It is also interesting to note that $\mmin/\md$ is independent
of the circular cylindrical radius $a$; it depends only upon 
the elongation $l/a$, as long as the needle is highly conducting 
so that $\varepsilon_0 \rightarrow \infty$.

\section{Discussion}\label{sec:discussion}
Metallic iron particles are expected to condense
in circumstellar shells irrespective of the O/C ratio
(Gilman 1969; Grossman 1972; Lewis \& Ney 1979;
Kozasa \& Hasegawa 1988). In O-rich stars, 
Gail \& Sedlmayr (1999) suggested that metallic iron
particles may form as inclusions embedded in large 
silicate grains.  
Kemper et al.\ (2002) argued
that nonspherical metallic particles can 
provide the 3--8$\mum$ opacity needed to 
fit the spectral energy distribution of 
the OH/IR star OH~127.8+0.0, suggesting 
that iron particles may be produced 
in quiescent O-rich stellar outflows.
Wickramasinghe \& Wickramasinghe (1993) modeled 
the IR emission of Supernova~1987A and argued for 
the condensation of metallic needles in its ejecta 
with an amount of $\simali$$8\times 10^{-8}\Msun$. 
As evolved stars and supernovae are the primary
sources of the dust in the ISM
(e.g., see Edmunds \& Morgan 2005, Gomez et al.\ 2007,
Dunne et al.\ 2009, Barlow et al.\ 2010, Matsuura et al.\ 2011),
it is quite plausible that iron grains may be an appreciable 
dust constituent of the general ISM.

The growth of metallic whiskers under laboratory conditions  
has been extensively demonstrated for a large number of elements 
such as carbon, iron, aluminum, magnesium, potassium and mercury.
Hoyle \& Wickramasinghe (1988) postulated that 
the growth of metallic whiskers in supernova ejecta begins as 
more or less spherical clusters of atoms with a radius that 
increases to $\simali$0.01$\mum$. Because of the appearance
of a helical or screw-dislocation which becomes 
self-propagating along the direction of the whisker,
the condensate grows linearly with great rapidity
up to lengths of $\simali$1\,mm or more. 
They further attributed the origin of the screw dislocation 
to an occasional radioactivity unstable cobalt atom that became 
incorporated in the initial more or less spherical growth of 
a condensate. According to Hoyle \& Wickramasinghe (1988), 
the rate of whisker growth is exponential;
for example, a needle of radius of 0.01$\mum$ could grow to
a length of $\simali$1$\mm$ in only $10^6\yr$ in a typical
protostellar environment. 
On the other hand, Piotrowski (1962)
argued that elongated grains, if electrically charged, 
would grow longer through preferential capture of ions 
near the ends of the grain. 
%
%{\bf 
In addition, there are also other processes that 
can produce chains of metallic 
(as well as insulating) particles 
in natural systems (e.g., the early solar nebula). 
Marshall et al.\ (2005) have experimentally shown 
that triboelectric charging of grains due to turbulent 
collisions followed by ``head-to-tail'' aggregation 
of the charged grains could lead to the formation
of filamentary aggregates at enhanced aggregation rates. 
Nuth et al.\ (1994) and Nuth \& Wilkinson (1995)
have also both experimentally and theoretically 
shown that the formation of chains of very small 
($\simali$20\,nm) iron grains in the solar nebula
could be greatly enhanced 
due to magnetic diploe interactions.
In both the electrostatic (Marshall et al.\ 2005) and
magnetic (Nuth et al.\ 1994, Nuth \& Wilkinson 1995)
aggregated cases, there can be an almost infinite length 
to diameter ratio.
%}

More recently, Nuth et al.\ (2010) have experimentally 
demonstrated the formation of abundant graphite whiskers 
on or from the surfaces of the graphite grains 
when they were repeatedly exposed to H$_2$, 
CO, and N$_2$ at 875$\K$, a condition mimicing 
that of protostellar nebulae. This naturally explains
the discovery of graphite whiskers in 
carbonaceous chondrites (Fries \& Steel 2008).
Nuth et al.\ (2010) further argued that graphite whiskers 
could be expelled from protostellar systems 
either in polar jets or by radiation pressure 
and thus populate the interstellar space
(also see Bland 2008).
%
%{\bf
The protostellar-nebula origin of metallic whiskers 
may be advantageous over
the circumstellar condensation scenario.
While vapor-phase growth along the c-axis of 
a crystalline metal grain is possible, 
many freshly condensed grains in circumstellar 
environments would be frozen ``liquid'' drops 
and require annealing prior to crystallization 
since crystallization would be required to yield 
a c-axis to direct further growth. 
However, once the initial condensation growth 
phase ceases, there is generally no more condensable 
material left in the gas phase to add to the crystal along 
any axial direction. This restriction does not necessarily 
apply in a protostellar nebula where large scale mixing 
can occur (see Nuth et al.\ 2010).
%}
%

In any case, to quantitatively evaluate the role 
of metallic needles in explaining the mid-IR
interstellar extinction, the dimming of Type Ia supernovae, 
the thermalization of the cosmic microwave background,
as well as the expulsion of metallic needles from
stellar and protostellar systems,
an accurate knowledge of the absorption cross sections
of metallic needles over a wide wavelength range is required.
We therefore call for rigorous calculations of the absorption
cross sections of metallic needle-like particles 
of various aspect ratios, 
presumably with the discrete dipole approximation 
(DDA; Purcell \& Pennypacker 1973, Draine 1988).
%{\bf
However, we should also note that in the IR, 
the complex dielectric functions of metallic materials 
are very large and the DDA often
does not give very accurate results.  
Also, cosmic needles could contain impurities, 
oxide coatings or other chemical heterogeneities. 
In addition, the needles found in astrophysical 
environments are not likely all single axis needles
(because there is no significant axial stiffness 
as would be found in a crystalline needle), 
but might consist of elongated grain aggregates 
with significant numbers of kinks and bends 
along the primary axis. 
We therefore also call for systematic
experimental explorations of the formation
and growth mechanisms of 
metallic needle-like particles 
in conditions which may prevail in protostellar nebulae
as well as experimental measurements 
of their absorption properties
over a wide wavelength range 
from the UV to the far-IR, 
submillimeter and millimeter.
%
%}
%

\section*{Acknowledgements}
We thank Drs. B.T.~Draine, E.~Dwek, F.Z.~Liu, 
D.~Pfenniger, and J.L.~Puget 
for very helpful discussions.
%{\bf
We are particularly indebted to Dr. J.A.~Nuth
whose stimulating comments and suggestions 
largely improved the presentation of this paper.
CYX is supported in part by 
the Talents Recruiting Program of Beijing Normal University
and the National Natural Science Foundation of China 
(NSFC) Grant No.\,91952111.
JHC is supported in part by 
NSFC Grant No.\,U1731107.
AL is supported in part by 
a NSF grant AST-1816411.
%and a NASA grant 80NSSC19K0575.
%}

\section*{Data Availability}
The data underlying this article will be shared 
on reasonable request to the corresponding authors.

\bsp
\label{lastpage}

\begin{thebibliography}{}
\bibitem[]{}Aguirre, A.N. 1999, ApJL, 512, L19
\bibitem[]{}Aguirre, A.N. 2000, ApJ, 533, 1
\bibitem[]{}Banerjee, S.K., Narlikar, J.V., Wickramasinghe, N.C.,
            Hoyle, F., \& Burbridge, G. 2000, AJ, 119, 2583
\bibitem[]{}Barlow, M.~J., Krause, O., Swinyard, B.~M., 
                  et al.\ 2010, A\&A, 518, L138
\bibitem[]{}Bennett, C.L., et al.\ 2003, ApJS, 148, 1
\bibitem[]{}Bland, P.A.\ 2008, Science, 320, 61
\bibitem[]{}Bradley, J.~P., Harvey, R.~P., \& McSween, H.~Y.\ 
                 1996, Geochim. Cosmochim. Acta, 60, 5149
\bibitem[]{}Chlewicki, G., \& Laureijs, R.J. 1988, A\&A, 207, L11
%\bibitem[]{}D\'{e}sert, F.X., Boulanger, F., \& Puget, J.L. 1990, 
%            A\&A, 237, 215
\bibitem[]{}Draine, B.T.\ 1988, ApJ, 333, 848
\bibitem[]{}Draine, B.T.\ 1989,
            in Infrared Spectroscopy in Astronomy,
            ed. B. H. Kaldeich
            (Paris: ESA Publ. Division), 93
%\bibitem[]{}Draine, B.T. 2003, ARA\&A, 41, 241
\bibitem[]{}Draine, B.T. 2003, in The Cold Universe,
            Saas-Fee Advanced Course Vol.\,32, 
            ed. D. Pfenniger (Berlin: Springer-Verlag), 213
%\bibitem[]{}Draine, B.T., \& Anderson, N. 1985, ApJ, 292, 494
\bibitem{}Draine, B.T., \& Hensley, B.\ 
                2013, ApJ, 765, 159 
\bibitem[]{}Draine, B.T., \& Li, A.\ 2001, ApJ, 551, 807
%\bibitem[]{}Draine, B.T., \& Li, A.\ 2007, ApJ, 657, 810
\bibitem[]{}Dunne, L., Eales, S., Ivison, R., Morgan, H.,
            \& Edmunds, M.\ 2003, Nature, 424, 285
\bibitem[]{}Dunne, L., Maddox, S.~J., Ivison, R.~J., et al.\
                  2009, MNRAS, 394, 1307
\bibitem[]{}Dwek, E. 2004a, ApJ, 611, L109
\bibitem[]{}Dwek, E. 2004b, ApJ, 607, 848
%\bibitem[]{}Dwek, E., et al.\ 1997, ApJ, 475, 565
\bibitem[]{}Edmunds, M.~G., \& Wickramasinghe, N.~C.\
                  1975, Nature, 256,713 
\bibitem[]{}Edmunds, M.~G., \& Morgan, H.~L.\ 2005, 
                  in The Fate of the Most Massive Stars
                  (ASP Conf. Ser., 332), 
                  ed. R. Humphreys \& K. Stanek 
                  (San Francisco, CA: ASP), 331
\bibitem[]{}Flaherty, K. M., Pipher, J. L.,
                  Megeath, S. T., et al.\ 2007, ApJ, 663, 1069
\bibitem[]{}Fries, M. \& Steele, A.\ 2008, Science, 320, 91
\bibitem[]{}Gail, H.-P., \& Sedlmayr, E. 1999, A\&A, 347, 594
%\bibitem[]{}Gall, C., Hjorth, J., Watson, D., et al.\ 
%                 2014, Nature, 511, 326
\bibitem[]{}Gao, J., Jiang, B.W., \& Li, A., 2009,
                  ApJ, 707, 89
\bibitem[]{}Gilman, R.C.\ 1969, ApJL, 155, L185
\bibitem[]{}Gomez, H.~L.\ 2014, Nature, 511, 296
\bibitem[]{}Gomez, H.~L., Dunne, L., Eales, S.~A., et al.\ 
                  2005, MNRAS, 361, 1012
\bibitem[]{}Gomez, H.~L., Eales, S.~A., \& Dunne, L.\ 
                  2007, Int. J. Astrobiol., 6, 159
\bibitem[]{}Gomez, H.~L., Dunne, L., Ivison, R.~J., et al.\ 
                  2009, MNRAS, 397, 1621
\bibitem[]{}Greenstein, J.L. 1938, Harvard Obs. Circ., No.\,422
\bibitem[]{}Grossman, L. 1972, Geochim. Cosmochim. Acta, 36, 597
\bibitem[]{}Indebetouw, R., Mathis, J. S., Babler, B. L., 
                  et al.\ 2005, ApJ, 619, 931
\bibitem[]{}Hensley, B.~S. \& Draine, B.~T.\ 
                  2017, ApJ, 834, 134
\bibitem[]{}Hensley, B.~S. \& Draine, B.~T.\ 
                  2020, ApJ, 895, 38
\bibitem[]{}Hoyle, F., \& Wickramasinghe, N.C. 
            1988, Ap\&SS, 147, 245
\bibitem{}Jenkins, E.~B.\ 2009, 
                ApJ, 700, 1299 
\bibitem[]{}Jiang, B.W., Gao, J., Omont, A.,
                  Schuller, F., \& Simon, G.\ 2006,
                  A\&A, 446, 551
%\bibitem[]{}Jones, A.P. 1990, MNRAS, 245, 331
\bibitem[]{}Kemper, F., de Koter, A., Waters, L.B.F.M.,
                 Bouwman, J., \& Tielens, A.G.G.M.\
                 2002, A\&A, 384, 585
\bibitem[]{}Krause, O., Birkmann, S.M., Rieke, G.H., 
                  et al.\ 2004, Nature, 432, 596
\bibitem[]{}Kozasa, T., \& Hasegawa, H.\ 1988, Icarus, 73, 180
\bibitem[]{}Lewis, J.S., \& Ney, E.P.\ 1979, ApJ, 234, 154
\bibitem[]{}Li, A.\ 2003a, ApJ, 584, 593
\bibitem[]{}Li, A.\ 2003b, ApJL, 599, L45
%\bibitem[]{}Li, A. 2004, in ASP Conf. Ser. 309,
%            Astrophysics of Dust, ed. A.N. Witt, 
%            G.C. Clayton, \& B.T. Draine 
%            (San Francisco: ASP), 417
\bibitem[]{}Li, A., \& Draine, B.T.\ 2001, ApJ, 554, 778
%\bibitem[]{}Li, A., \& Draine, B.T.\ 2002, ApJ, 576, 762
%\bibitem[]{}Li, A., \& Draine, B.T.\ 2012, ApJL, 760, L35
%\bibitem[]{}Li, A., \& Greenberg, J.M. 2003, in Solid State 
%            Astrochemistry, ed. V. Pirronello, J. Krelowski, 
%            \& G. Manic\'o (Dordrecht: Kluwer), 37
%\bibitem[]{}Lutz, D., et al. 1996, A\&A, 315, L269
\bibitem[]{}Lutz, D. 1999,
            %ISO Observations of the Galactic Centre,
            in The Universe as Seen by ISO,
            ed. P. Cox \& M. Kessler
            (ESA Special Publ., Vol.~427;
             Noordwijk: ESA), 623
%{\bf
\bibitem[]{}Marshall, J.~R., Sauke, T.~B., 
                   \& Cuzzi, J.~N.\ 2005,
                   Geophys. Res. Lett., 32, L11202
%}
\bibitem[]{}Matsumoto, T., Harries, D., Langenhorst, F., 
                  et al.\ 2020, Nature Comm., 11, 1117
\bibitem[]{}Matsuura, M., Dwek, E., Meixner, M., et al.\ 
                  2011, Science, 333, 1258
\bibitem[]{}Melia, F.\ 2020, European Phys. J. Plus, 135, 511
\bibitem[]{}Morgan, H.~L., Dunne, L., Eales, S.~A., 
                 et al.\ 2003, ApJL, 597, L33
\bibitem[]{}Narlikar, J.V., Vishwakarma, R.G., Hajian, A.,
            Souradeep, T., Burbridge, G., \& Hoyle, F. 2003, 
            ApJ, 585, 1 
\bibitem[]{}Nishiyama, S., Tamura, M., Hatano, H.,
                 et al.\ 2009, ApJ, 696, 1407
%{\bf
\bibitem[]{}Nuth, J.~A. \& Wilkinson, G.~M.\ 
                    1995, Icarus, 117, 431
\bibitem[]{}Nuth, J.~A., Berg, O., Faris, J., 
                    \& Wasilewski, P.\
                   1994, Icarus, 107, 155
%}
\bibitem[]{}Nuth, J.~A., Kimura, Y., Lucas, C., et al.\ 
                  2010, ApJL, 710, L98
%\bibitem[]{}Oort, J.H., \& van de Hulst, H.C. 1946,
%            Bull. Astron. Inst. Netherlands, 10, 187
\bibitem[]{}Perlmutter, S., et al.\ 1999, ApJ, 517, 565
\bibitem[]{}Piotrowski, S.L. 1962, Acta Astron., 12, 221 
\bibitem[]{}Poteet, C.~A., Whittet, D.~C.~B., 
                  \& Draine, B.~T.\ 2015, ApJ, 801, 110
\bibitem[]{}Purcell, E.M. 1969, ApJ, 158, 433
\bibitem[]{}Purcell, E.~M., \& Pennypacker, C.~R.\ 
                 1973, ApJ, 186, 705
\bibitem[]{}Riess, A. G., et al.\ 1998, AJ, 116, 1009
\bibitem[]{}Schal\'{e}n, C. 1936, Medd. Uppsala Astron. Obs., No.\,64
%\bibitem[]{}Siebenmorgen, R., \& Kr\"{u}gel, E. 1992, A\&A, 259, 614
\bibitem[]{}Wang, S., Gao, J., Jiang, B.W., Li, A.,
                 \& Chen, Y.\ 2013, ApJ, 773, 30
\bibitem[]{}Wang, S., Li, A., \& Jiang, B.W.\ 2014,
                  Planet. Space Sci., 100, 32
\bibitem[]{}Wang, S., Li, A., \& Jiang, B.W.\ 2015a,
                  ApJ, 811, 38
\bibitem[]{}Wang, S., Li, A., \& Jiang, B.W.\ 2015b,
                  ApJ, 454, 569
\bibitem[]{}Weingartner, J.C., \& Draine, B.T. 2001, ApJ, 548, 296
%\bibitem[]{}Weiland, J.L., Blitz, L., Dwek, E., Hauser, M.G., 
%            Magnani, L., \& Rickard, L.J. 1986, ApJ, 306, L101
\bibitem[]{}Westphal, A.~J., Butterworth, A.~L., 
                  Tomsick, J.~A., et al.\ 2019, ApJ 872, 66
%\bibitem[]{}Wickramasinghe, N.C. 1992, Ap\&SS, 198, 161
\bibitem[]{}Wickramasinghe, N.C., \& Wickramasinghe, A.N.\
            1993, Ap\&SS, 200, 145
\bibitem[]{}Wilson, T.~L., \& Batrla, W.\ 2005, A\&A, 430, 561
\bibitem[]{}Wright, E.L.\ 1982, ApJ, 255, 401
\bibitem[]{}Xue, M., Jiang, B.W., Gao, J., et al.\
                  2016, ApJS, 224, 23
%\bibitem[]{}Zubko, V.G., Dwek, E., \& Arendt, R.G. 
%          2004, ApJS, 152, 211
\end{thebibliography}
\end{document}